\documentclass[aps,prb,reprint,superscriptaddress]{revtex4-2}

\usepackage{tikz}
\usepackage{scalerel}
\usepackage{amssymb}
\usepackage{suffix}
\usepackage{float}
\usepackage{mathtools}
\usepackage[utf8]{inputenc}
\usepackage{booktabs}
\usepackage{cases}
\usepackage[multiple]{footmisc}
\usepackage{physics}
\usepackage{dcolumn}
\usepackage{color,soul}
\usepackage{rotating}
\usepackage{perpage}
\usepackage{xcolor}
\usepackage{comment}
\usepackage{soul}
\usepackage{tikz}
\usepackage[T1]{fontenc}
\usepackage{etoolbox}
\usepackage{graphics}
\usepackage{siunitx}
\usepackage{hyperref}
\hypersetup{colorlinks}
\usepackage{float}	
\usepackage{collref}
\usepackage{multirow}
\usepackage{mathtools}
\usepackage{bm}
\usepackage{url}
\usepackage{wasysym}

\begin{document}
\title{Coherent matter wave emission from an atomtronic transistor}
\author{Sasanka Dowarah}
\email{Sasanka.Dowarah@utdallas.edu}
 \affiliation{Department of Physics, The University of Texas at Dallas, Richardson, Texas 75080, USA}
\author{Mengxin Du}
 \affiliation{Department of Physics, The University of Texas at Dallas, Richardson, Texas 75080, USA}
\author{Alan Zanders}
 \affiliation{Department of Physics, The University of Texas at Dallas, Richardson, Texas 75080, USA}
 
 \author{Shengwang Du}
 \affiliation{Department of Physics, The University of Texas at Dallas, Richardson, Texas 75080, USA}
 \affiliation{
Elmore Family School of Electrical and Computer Engineering, Purdue University, West Lafayette, Indiana 47907, USA}
\affiliation{Department of Physics and Astronomy, Purdue University, West Lafayette, Indiana 47907, USA}
\affiliation{Purdue Quantum Science and Engineering Institute, Purdue University, West Lafayette, Indiana 47907, USA}

\author{Michael Kolodrubetz}
\email{mkolodru@utdallas.edu}
 \affiliation{Department of Physics, The University of Texas at Dallas, Richardson, Texas 75080, USA}

 \author{Chuanwei Zhang}
\email{Chuanwei.zhang@wustl.edu}
 \affiliation{Department of Physics, The University of Texas at Dallas, Richardson, Texas 75080, USA}
 \affiliation{Department of Physics, Washington University, St. Louis, MO, 63130, USA}
 
\date{\today}
\begin{abstract}
The atomtronic matter-wave triple-well transistor is theoretically predicted to exhibit current gain and act as a coherent matter-wave emitter. In this work, we investigate the dynamics of an atomtronic transistor composed of a triple-well potential -- source, gate, and drain -- modeled by the time-dependent Gross-Pitaevskii equation. We systematically explore the dependence of the drain population and the current on the source bias potential and the strength of the interatomic interaction. Our simulations reveal signatures of resonant tunneling when the source chemical potential aligns with discrete energy levels in the gate well, leading to coherent matter-wave emission in the drain. Contrary to previous many-body studies that predicted interaction-induced current gain via coupling to gate well modes, our results suggest that coherence in the drain is primarily governed by single-particle resonances, with no evident broadening from nonlinear coupling.
\end{abstract}             
\maketitle
{\allowdisplaybreaks}
 \parskip 0 pt
\section{Introduction}
Atomtronics is a paradigm that uses ultracold atoms for coherent manipulation of matter and information processing \cite{RevModPhys.94.041001, Amico2021Roadmap}. Various atomtronics components such as batteries \cite{PhysRevA.88.043641}, diodes \cite{PhysRevA.75.023615, PhysRevLett.103.140405}, and transistors \cite{PhysRevA.104.033311, PhysRevLett.103.140405} have been constructed in analogy to their electronic counterparts, with charge flow replaced by the flow of atoms. While inspired by electronics, atomtronics devices exhibit different properties than the former. For instance, the atomtronic battery can exhibit both positive and negative internal resistance  \cite{PhysRevA.88.043641}. 

\par
\begin{figure*}[!]
\includegraphics[width=\linewidth]{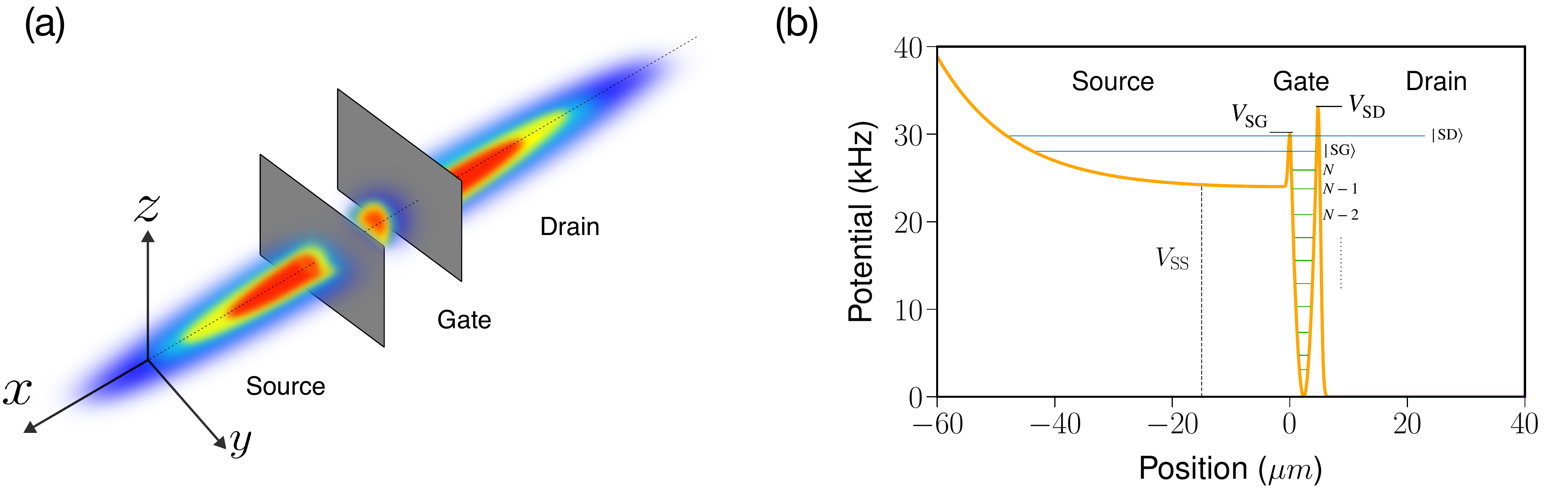}
\caption{\label{fig:schematic_experimental_setup}(a) A 3-D schematic of the experimental realization of the atomtronic transistor used in experiments. An elongated, cigar-shaped Bose-Einstein Condensate (BEC) is held in a highly anisotropic trap, with its longest dimension along the $x-$axis. Two repulsive optical barriers partition the condensate into the source, gate, and drain regions. Due to high confinement along the radial direction ($y$ and $z$ axes), the dynamics effectively take place only along the $x$-axis. (b) The one-dimensional, triple-well potential landscape of the atomtronic transistor with the ``source'' well that supplies the atoms, the nearly harmonic ``gate'' well in the middle, and the flat ``drain'' well. The drain well extends far enough to prevent atoms from reflecting. The barrier heights are $V_{\mathrm{SG}} = 31$ kHz (between source and the gate) and $V_{\mathrm{GD}} = 33$ kHz (between gate and the drain), respectively. The gate well spans from $x_{\mathrm{SG}}=0$ to $x_{\mathrm{GD}}=4.8\;\mu$m, which we choose to be consistent with the experiments \cite{Caliga_2016_principle, Caliga_2016}. These structural parameters of the potential are held constant throughout all simulations.}
\end{figure*}

 The atomtronic transistor consists of a triple-well potential through which ultracold atoms -- generally a Bose-Einstein condensate (BEC) flow -- these three wells are referred to as the -- ``source  (S)'', ``gate (G)'' and the ``drain (D)'' well as in the electronic field effect transistor (illustrated in Fig. \ref{fig:schematic_experimental_setup}) \cite{PhysRevA.104.033311}. A bias potential $V_{\mathrm{SS}}$ is applied to the source well that is used to control the flow of atoms from the source well. The ensemble of atoms is initially placed in the source well with a chemical potential of $\mu_{B}$ and at a temperature of $T_{B}$. It powers the circuit, and therefore referred to as the `battery' \cite{PhysRevA.104.033311}. The atom current then flows to the gate and the drain well, which are initially empty. Experimentally, it was found that by controlling the relative heights of the two barriers -- source-gate $V_{\mathrm{SG}}$, and the gate-drain $V_{\mathrm{GD}}$, the transistor dynamics can be controlled  \cite{Caliga_2016}. For specific ranges of these parameters, a BEC can form in the gate well, and the gate's chemical potential can exceed that of the source, which has been explained by a semiclassical treatment \cite{Caliga_2016_principle}.

An approximate many-body theoretical framework has been developed in  \cite{PhysRevA.104.033311} to explain the mechanisms of current gain and the emission of a coherent matter wave from the transistor. This theory models the gate as a harmonic well with a finite number of bound states $\{|n\rangle, n=0,1,\cdots, N\}$ which is assumed to have negligible tunneling to the source and the drain well.  Additionally, the theory assumes that two additional energy levels $|N+1\rangle \equiv  |\mathrm{SG}\rangle, |N+2\rangle \equiv |\mathrm{SD}\rangle$ ( referred to as the transistor modes), couple the source to the gate, and the source to the drain well respectively (see Figure \ref{fig:schematic_experimental_setup}). According to this model, a condensate forms in the gate well as a coherent state (or a displaced ground state) and oscillates at the single-particle gate frequency. This oscillating condensate couples to the transistor modes, which broadens the resonant tunneling window for incoming atoms. The result is an enhanced drain current, a phenomenon termed ``current gain.'' A key prediction of this model is that the emitted matter wave in the gate well should oscillate at the single particle oscillation frequency of the gate well, $\omega_{\mathrm{Gate}}$, rather than the frequency determined by the energy of these particles $\omega_{D}=E_{\mathrm{SD}}/\hbar$, where $E_{\mathrm{SD}}$ is the energy corresponding to the transistor mode $|\mathrm{SD}\rangle = |N + 2 \rangle$.

In this work, we numerically investigate the prediction of coherent matter-wave emission from an atomtronic transistor using the time-dependent Gross-Pitaevskii equation (GPE). We systematically explore how the coherence of the emitted matter wave and the transport of atoms are affected by key system parameters, with a particular focus on the role of interatomic interaction strength. Our analysis is conducted within the mean-field, zero-temperature framework of the Gross-Pitaevskii equation (GPE), which does not account for thermal effects. Furthermore, all simulations are performed in the regime where the source-gate barrier is lower than the gate-drain barrier $(V_{\mathrm{SG}} < V_{\mathrm{GD}})$, consequently, the influence of the feedback parameter on the transistor dynamics is not addressed in this study.
\section{Model}
We model the atomtronic transistor based on a triple-well potential structure used in experiments \cite{Caliga_2016, Caliga_2016_principle}. A schematic of the physical realization of this system is shown in Figure \ref{fig:schematic_experimental_setup}(a). The setup consists of a Bose-Einstein Condensate (BEC) confined in a highly anisotropic, ``cigar-shaped'' trap, with its longest dimension oriented along the $x-$axis. Two repulsive optical barriers are shown cutting the atom cloud, which partitions the condensate into the source, gate, and the drain well, as labeled in figure \ref{fig:schematic_experimental_setup}(a). The strong confinement along the $y,z$ directions restricts the dynamics primarily along the longitudinal $x-$axis, justifying the use of a one-dimensional model for theoretical analysis, as will be described in the next section.
\subsection{The Gross-Pitaevskii equation}
The Gross-Pitaevskii equation \cite{Gross1961, Pitaevskii1961} describes BEC dynamics in the mean-field approximation at temperatures near absolute zero. It describes the ground state of many bosons using a wavefunction by approximating the condensate as a self-interacting wave function. In the atomtronics experiments, the atom ensemble consists of both thermal $N_{\mathrm{th}}$, and condensed atoms $N_{c}$. Even though both the thermal and condensed atoms take parts in the transistor dynamics, we will work in the limit $N_{c} \gg N_{\mathrm{th}}$, where the Gross-Pitaevskii equation has been used to predict qualitative behaviors of the dynamics \cite{RevModPhys.71.463}. We choose parameters to align with recent atomtronic transistor experiments \cite{Caliga_2016, caliga2013matterwavetransistoroscillator}, namely the longitudinal gate trap frequency, $\omega_{\mathrm{Gate}, l}$, and with the transverse trap frequency $\omega_{r} = 10\omega_{  \mathrm{Gate}, l}$, such that the transistor dynamics effectively takes place along the longitudinal direction only. For such cigar shaped clouds with $\omega_{r} \gg \omega_{l}$, the dynamics of Bose-Einstein condensate can be described by the one-dimensional Gross-Pitaevskii equation (GPE) \cite{PhysRevA.83.043611, Pigneur2020, PhysRevA.70.023604, PhysRevA.65.043614}
\begin{multline*}
i\hbar \pdv{\psi(x,t)}{t} = -\frac{\hbar^{2}}{2m} \pdv[2]{\psi(x,t)}{x} + V_{\mathrm{ext}}(x,t)\psi(x,t)\\ 
+ 2\hbar \omega_{r} a_{s} N_{\mathrm{total}} |\psi(x,t)|^{2}\psi(x,t),
\end{multline*}
with the single particle wavefunction normalized to unity:
\begin{eqnarray}
    \int dx |\psi(x,t)|^2 &=& 1.
\end{eqnarray}
Here $N_{\mathrm{total}}$ is the total number of atoms in the trap, $m$ is the atom mass, $V_{\mathrm{ext}}(x)$ is the external transistor potential, and $a_{s}$ is the $s-$wave scattering length for the atoms.
For numerical efficiency we introduce the following dimensionless variables
\begin{eqnarray}
    \Tilde{x} &=& x/l_{0},
    \;\Tilde{t}=t/t_{0},\;
    \Tilde{\psi}=\psi\sqrt{l_{0}}
\end{eqnarray}
with $l_{0} = \sqrt{\hbar/(m\omega_{l})}$, $t_{0} = 1/\omega_{l}$ where $\omega_{l} = \mathrm{max}(\omega_{l,\mathrm{source}}, \omega_{l,\mathrm{gate}}, \omega_{l,\mathrm{drain}})$. Using these parameters, we obtain the dimensionless GPE
\begin{eqnarray}
   i\pdv{\Tilde{\psi}}{\Tilde{t}} &=&  -\frac{1}{2}\pdv[2]{\Tilde{\psi}}{\Tilde{x}} + \Tilde{V}\Tilde{\psi} + \Tilde{g}|\Tilde{\psi}|^{2}\Tilde{\psi}, \label{eq:dimensionless-GP}
\end{eqnarray}
where $\Tilde{V} = V/(\hbar \omega_{l})$ is the dimensionless external transistor potential and $\Tilde{g} = 2\omega_{r}a_{s}N_{\mathrm{total}}/(\omega_{l}l_{0})$ is the dimensionless interaction strength. Other parameters include: the mass of the Rubidium$-87$ atom, $M_{\mathrm{Rb}} = 1.419\times 10^{-25}$ kg, $s-$wave scattering length, $a_{s} = 5.186\times 10^{-9} $ m  \cite{PhysRevA.87.053614}, number of atoms $N_{\mathrm{total}} = 10^{4}$. With these parameters, the dimensionless interaction parameter has the value
\begin{eqnarray}
    \Tilde{g} &=& 3273.24\quad .\label{eq:dimensionless_interaction}
\end{eqnarray}

The transistor experiment starts with a BEC in the source well with high confining potential barriers on both sides  \cite{Caliga_2016}. To emulate this numerically, we solve for the ground state of Eq.\eqref{eq:dimensionless-GP} in the source well with high initial barriers placed at both ends of the source well, using imaginary time evolution \cite{PhysRevE.62.7438,LEHTOVAARA2007148, 10.10635.0143556}, and Split-step Fourier method \cite{doi:10.1137/0705041, TAHA1984203, Bao-2013}. After obtaining the ground state, we lower the barriers to the transistor setting, and then time evolve the wavefunction using the GPE [Eq.\eqref{eq:dimensionless-GP}]. To test numerical convergence and stability, we perform the simulation with the position grid divided into $N=2^{15} \; ( \Delta x = 9.46\times 10^{-8}$ m) and $2^{16}\; (\Delta x = 4.73\times 10^{-8}$ m) divisions, and the time grid into intervals of $\Delta t=10^{-6}$ s, $\Delta t=10^{-7}$ s and found that both of these give the same results for all relevant quantities. We therefore carry out the rest of the results with $N=2^{15}$ and $\Delta t=10^{-6}$ s. With the position being discretized, the number of atoms between the points $x=x_{1}$ and $x=x_{2}$ at time $t$ is calculated using the sum
\begin{eqnarray}
    N_{\mathrm{atoms}}(x_{1},x_{2},t)
    &\approx & N_{\mathrm{total}} \sum^{x_{2}}_{x=x_{1}} |\psi(x,t)|^{2} \Delta x. \label{eq:number_of_atoms_calculation}
\end{eqnarray}\par The tunneling of atoms from the source well to the gate well depends on the chemical potential of the atomic BEC. Since the system is in the limit where interaction energy is very large compared to the kinetic energy $ (\mathrm{K.E/P.E}\sim 10^{-4})$, the chemical potential of the atoms in the source well can be calculated in the Thomas-Fermi limit as  \cite{pitaevskii2003bose}
\begin{eqnarray}
    V_{\mathrm{ext}}(x) + 2\hbar \omega_{r}a_{s}N_{\mathrm{total}}|\psi(x)|^{2} &=& \mu.\label{eq:chemical-potential-in-TF-limit}
\end{eqnarray}
From Eq.\eqref{eq:chemical-potential-in-TF-limit} we see that once the external potential of the transistor is fixed, two independent parameters will affect the transistor dynamics -- the interaction strength, which is characterized by the scattering length $a_{s}$, and the bias potential in the source well $V_{\mathrm{SS}}$. Increasing the scattering length $a_{s}$ leads to an increase in the interaction between the atoms. On the other hand, increase (decrease) in the source bias potential $V_{\mathrm{SS}}$ leads to the increase (decrease) in the chemical potential of the atoms in the source well. These two quantities can also potentially be controlled in experiments. We will discuss the effect of these two independent parameters on the flow of atoms and the coherence of the matter wave.\par The middle well (gate) plays a significant role in controlling the coherence of the matter wave. Previous theoretical work has modeled the gate as a simple harmonic potential \cite{PhysRevA.104.033311}. In our analysis, we do not directly use this harmonic approximation. However, it is useful to expand around the minimum of the well to obtain an approximate single-particle oscillation frequency along the longitudinal direction as
\begin{eqnarray}
\omega_{\mathrm{Gate}, l} \equiv  \omega_{\mathrm{Gate}} &\approx& 
\sqrt{\frac{8 V_{\mathrm{SG}}}{M_{\mathrm{Rb}}(x_{\mathrm{GD}}-x_{\mathrm{SG}})^{2}}}.
\end{eqnarray}
Note that since the potential is a truncated version of the ideal harmonic well, this approximation becomes increasingly inaccurate especially near the edges around the top of the source-gate and the gate-drain barriers. Substituting the parameters from the transistor in figure \ref{fig:schematic_experimental_setup} yields $\omega_{\mathrm{Gate}} \approx 6973$ rad/s, which corresponds to an oscillation period of $T_{\mathrm{Gate}} \approx 0.85$ millisecond. The gate energy levels are separated by an energy gap of approximately $\hbar \omega_{\mathrm{Gate}}$. In the context of the transistor's operation, where atoms tunnel from the source to the drain via the gate, we should expect the atom current in the drain to exhibit a strong temporal coherence, oscillating at a frequency near $\omega_{\mathrm{Gate}}$. A Fourier analysis of the time-dependent atom density in the drain well may then reveal a dominant peak around $\omega_{\mathrm{Gate}}$, which would support our hypothesis that BEC dynamics in the gate region govern the coherence of the emitted matter wave in the drain.
\section{Results}
\subsection{Effect of source bias potential on transistor dynamics}
\begin{figure}
    \centering
    \includegraphics[width=\linewidth]{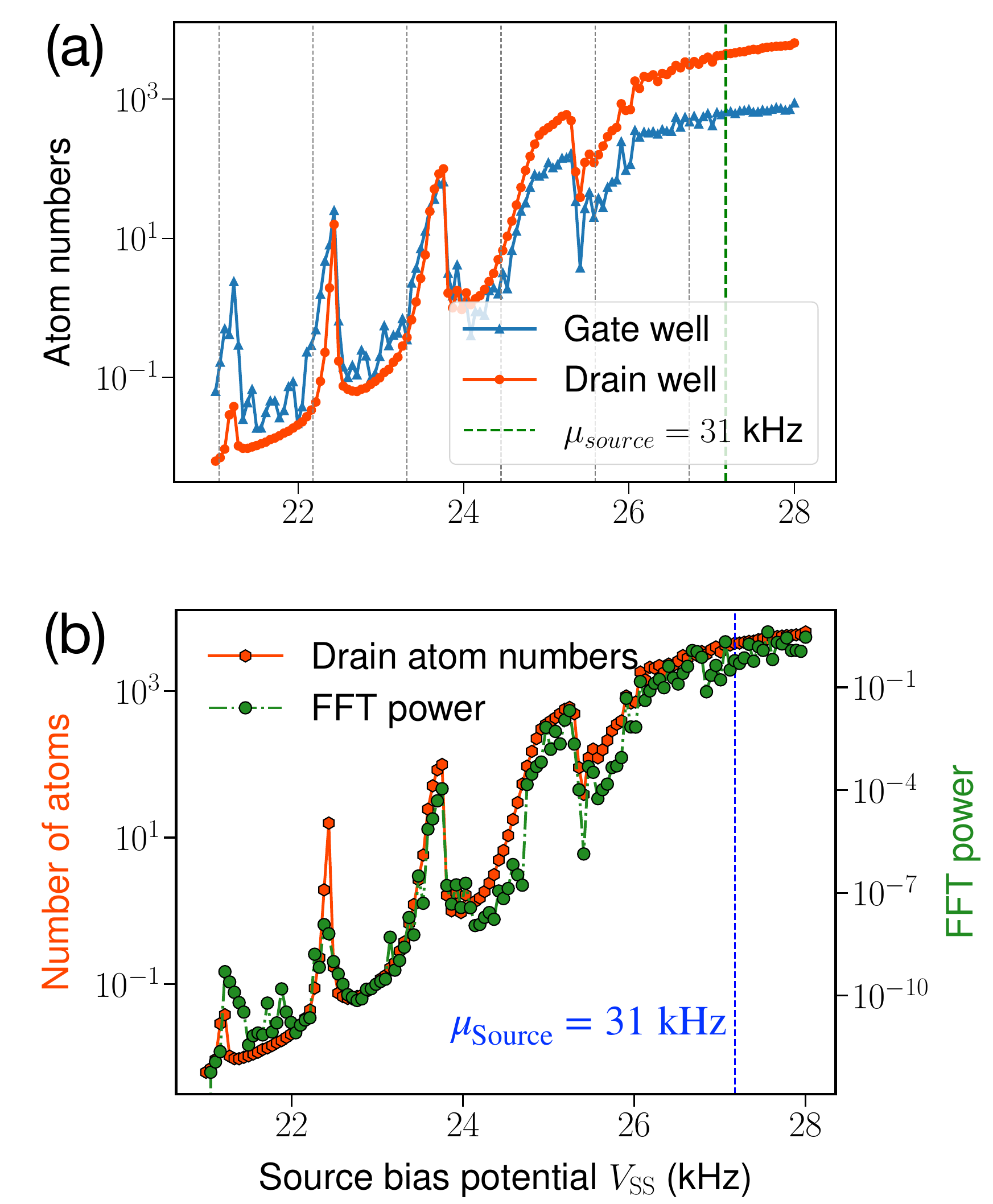}
    \caption{ (a) Number of atoms in the gate and the drain well after $t=200$ ms for $a_{s}=5.186\times10^{-10}$m. The gray vertical dotted lines show the single particle energy levels in the gate well. The green vertical dotted line indicates the value of the source bias potential at which the chemical potential of the atoms in the source well at $t=0$ exceeds the source gate barrier $V_{\mathrm{SG}}$, after which the dynamics is no longer only tunneling. (b) Drain atom population and signal coherence versus source bias potential. The power spectrum is calculated for the signal $|\psi(x =40 \; \mu m, t)|^{2}$ from $t= 0$ to $t= 300$ ms for $a_{s} = 5.186 \times 10^{-10}$ m. The peaks in both quantities are strongly correlated, indicating that maximum coherence occurs at resonant tunneling peak. The vertical blue dotted line marks the threshold where the
initial source chemical potential exceeds the source-gate barrier, $V_{\mathrm{SG}}$.}
    \label{fig:total_number_of_atoms_at_t=200ms_FFT}
\end{figure}

The chemical potential of the atoms in the source well can be tuned by adjusting the source bias potential. For instance, an ensemble of atoms that initially exhibits negligible tunneling because of a low chemical potential can be induced to tunnel by increasing the bias potential.  To illustrate this effect in our simulations, we computed and plotted, in Figure \ref{fig:total_number_of_atoms_at_t=200ms_FFT}(a), the total populations in both the gate and the drain wells at $t=200$ ms as a function of the applied bias potential, while maintaining the strength of the interatomic interaction constant. The atoms show periodic spikes in tunneling as the bias potential is increased. The resonant tunneling occurs approximately when the source chemical potential aligns with one of the discrete eigenenergies of the gate well. Since the drain well is essentially a continuum of energy levels, once the atoms resonantly tunnel to the gate well, they tunnel to the drain well freely. This behavior underscores the critical role of chemical potential matching in enabling and controlling coherent tunneling through the atomtronic transistor.
\subsection{Coherence of matter wave in the drain well}
\begin{figure}[h!]
    \centering
    \includegraphics[width=0.45\textwidth]{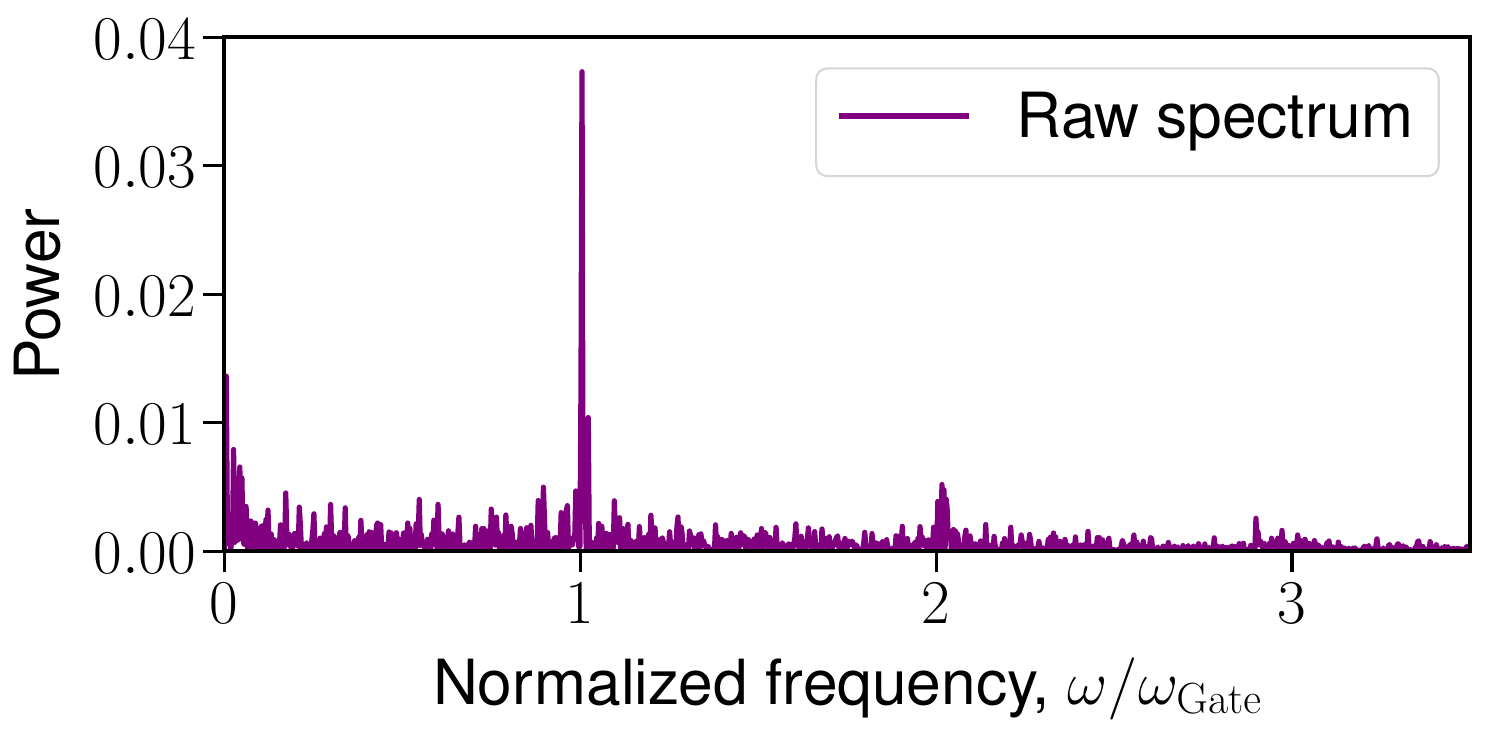}
    \includegraphics[width=0.45\textwidth]{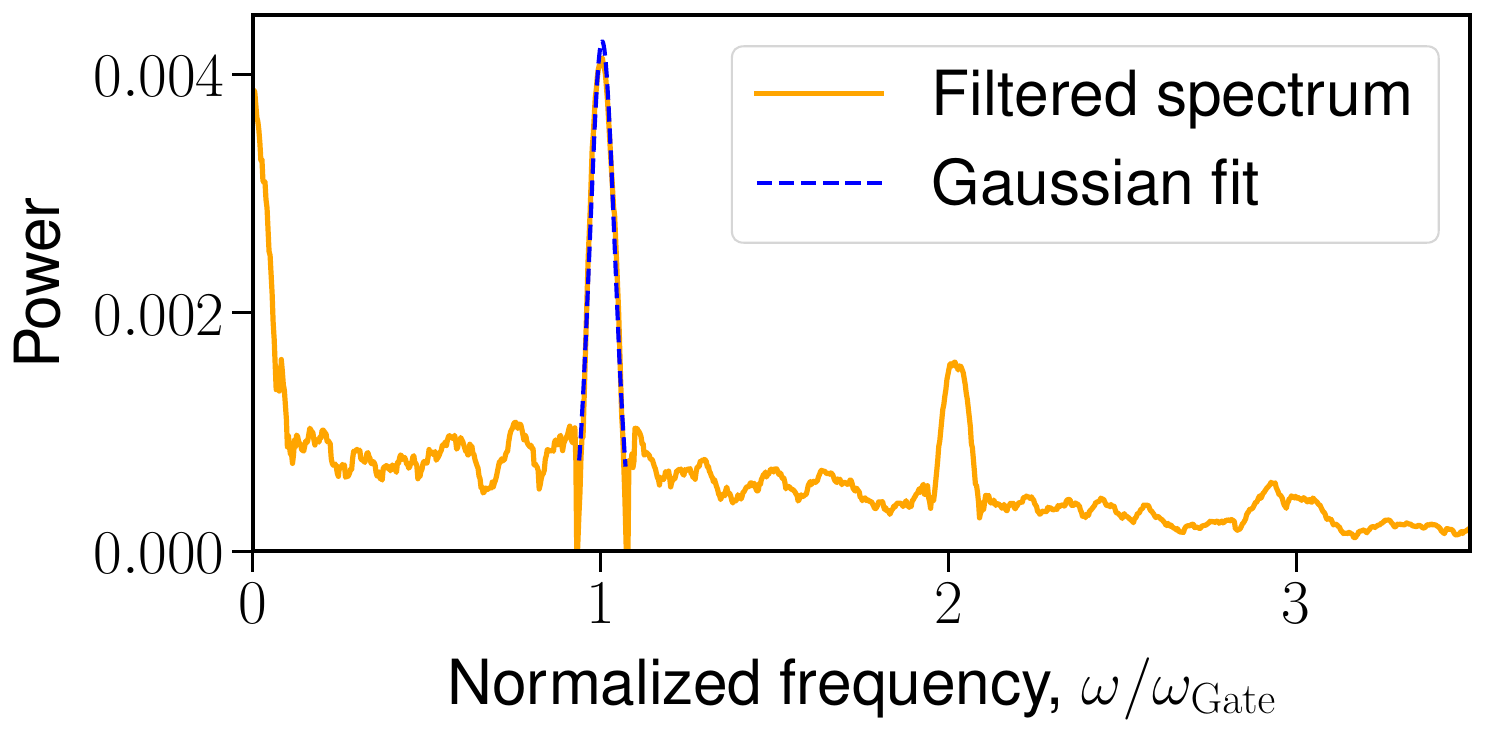}
    \caption{(Top) Raw power spectrum of the atom density $n(t) = |\psi(x=40\mu m, t)|^{2}$ from $t=0$ to $t=300$ ms for $a_{s} = 0.5\times 10^{-9}$m and $V_{\mathrm{SS}} = 26.4 $ kHz. (Bottom) The same power spectrum with Savitzky-Golay filter (window width $51 (1068.14$ rad/s), and polynomial order $2$), showing a Gaussian fit to the peak at $\omega=\omega_{\mathrm{Gate}}$.}
    \label{fig:raw_spectrum_and_filtered_spectrum}
\end{figure}

To test the coherent matter wave emission using our simulation, we look at the dominant frequencies in the atom density $n(t)\equiv |\psi(x_{0},t)|^{2}$, where $x_{0}$ is a fixed point in the drain well. The coherence in the matter wave in the drain well will be indicated by a peak in the power spectrum near $\omega=\omega_{\mathrm{Gate}}$.
As shown in Figure \ref{fig:raw_spectrum_and_filtered_spectrum}, in its raw form, the power spectrum is too noisy for analysis. We therefore use a Savitzky-Golay filter with a window size of $51$ corresponding to a window of size $1068.14$ rad/s in angular frequency units, and a polynomial order of $2$ to smooth the signal before extracting the peak characteristics. The analysis of the filtered spectrum confirms the presence of a coherent oscillation, revealing a prominent peak near $\omega=\omega_{\mathrm{Gate}}$. First, we examine how this peak in the power spectrum varies as a function of source bias potential for a fixed interaction strength. As shown in Figure \ref{fig:total_number_of_atoms_at_t=200ms_FFT}(b) the peaks in the power spectrum coincides with the peaks in the atom number in the drain well of the transistor. It demonstrates that the coherence of the emitted matter wave is intrinsically linked to the mechanism of resonant tunneling. Since the maximum atom transport occurs during resonant tunneling, the alignment of these peaks indicates that coherence is mainly coming from the single particle resonance effect in the gate well.  When the system is off-resonance, both the atom transport and the corresponding coherence signal are suppressed.

To quantify this coherence and its dependence on the interaction strength, we follow the following steps
\begin{enumerate}
    \item We begin by fixing the interaction strength between the atoms by choosing a scattering length $a_{s}$.
    \item For this value of $a_{s}$, we select a range of source bias potential $V_{\mathrm{SS}}$, ensuring that the initial chemical potential of the source‐well condensate lies below the height of the source–gate barrier $V_{\mathrm{SG}}$, ensuring that subsequent dynamics are dominated by quantum tunneling.
    \item For each source bias potential in the selected range, we time-evolve the wavefunction using the GPE. During each simulation, we record $\psi(x, t)$ in the drain well at $x = 40 \; \mu m$, in accordance with the experiments \cite{12479}, and compute the atom density signal $n(t)=|\psi(x = 40 \; \mu m, t)|^{2}$  from $t=0$ to $t=300$ ms.
    \item We then perform a Fourier transform of $n(t)$ to obtain the power spectrum and to determine the dominant frequencies in the signal.
    \item We observe a peak in the power spectrum approximately at the single particle gate frequency $\omega_{\mathrm{Gate}}$ as illustrated in Figure \ref{fig:raw_spectrum_and_filtered_spectrum}. We then fit the power spectrum data inside the interval $(\omega_{\mathrm{Gate}}/2, 3\omega_{\mathrm{Gate}}/2)$ with a Gaussian function to extract the peak. To map out the maximum peak/coherence, we vary the source bias potential and record the maximum peak amplitude in the power spectrum.
    \item Finally, we repeat this entire procedure for a series of scattering lengths 
    $a_{s}$ to assess the influence of interaction strength on the drain‐wave coherence. This procedure is necessary to compare different values of $V_{\mathrm{SS}}$ and $a_{s}$, subtracting off trivial effects such as exponential dependence of the tunneling rate on chemical potential.
\end{enumerate}

\begin{figure}[h!]
    \centering
    \includegraphics[width=0.45\textwidth]{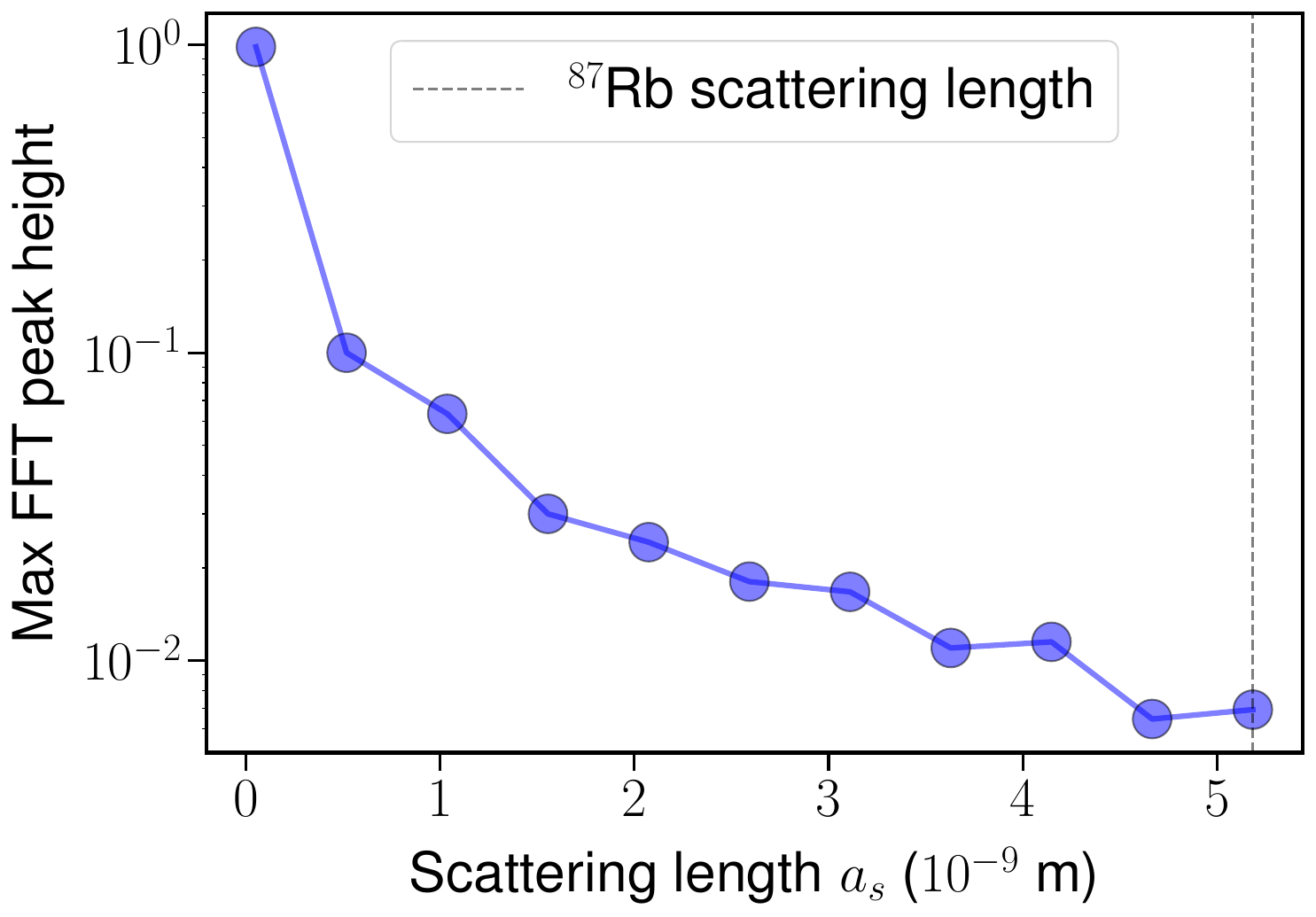}
    \caption{Maximum FFT power in the frequency window $\omega=(\omega_{\mathrm{Gate}}/2, 3\omega_{\mathrm{Gate}}/2)$ for all $V_{\mathrm{SS}} < \mu_{\mathrm{source}}$ values plotted as a function of the scattering length $a_{s}$.}
\label{fig:FFT_power_as_function_of_V_SS_different_a_s}
\end{figure}
In Figure \ref{fig:FFT_power_as_function_of_V_SS_different_a_s}, we plot the maximum FFT power found across all scanned source bias potentials in the source well for each given scattering length $a_{s}$. The results show a clear and monotonic trend: the maximum attainable degree of coherence decreases as the interaction strength increases. This suggests that strong interatomic interactions degrade the matter wave's coherence and suppress tunneling to the drain. 
\begin{figure}[h!]
    \centering
    \includegraphics[width=0.9\linewidth]{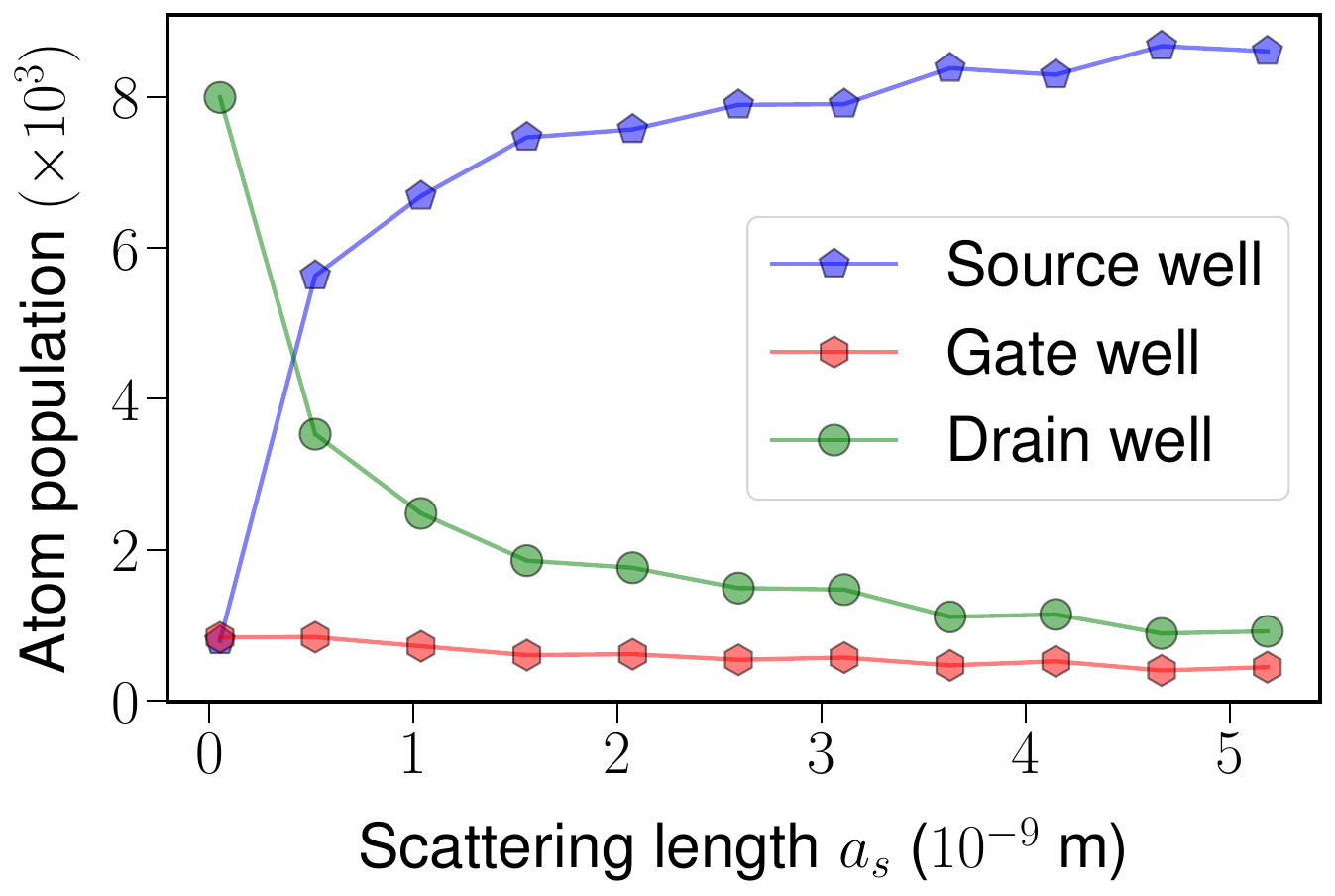}
    \caption{Atom populations as a function of scattering length $a_{s}$ calculated at the source bias values that give
  the maximum coherence. The number of atoms in the source, gate, and drain wells is recorded at $t=300$ ms. For each $a_{s}$, the data corresponds to the source bias $V_{\mathrm{SS}}$
  that yields the maximum FFT power of the drain signal.}
    \label{fig:atom_numbers_as_function_of_a_s}
\end{figure}\par
Figure \ref{fig:atom_numbers_as_function_of_a_s} illustrates the effect of the interaction strength on the final distribution of atoms across the triple-well potential, recorded at the source bias that yields maximum coherence. The plot shows that only a small fraction of atoms tunnel to the drain well within the $t=300$ ms time. The populations in the drain well exhibit a strong and monotonic decrease as the scattering length, $a_{s}$ increases, leading to a maximum drain population in the non-interacting limit $(a_{s}\to 0)$. The increase in the interaction strength also leads to the suppression of tunneling of atoms to the drain well, which also shows that by tuning the interaction strength between the atoms, we can control the drain well atom population.

Our results in Figure \ref{fig:FFT_power_as_function_of_V_SS_different_a_s} and Figure \ref{fig:atom_numbers_as_function_of_a_s} present insights into the transistor operation within the mean-field limit framework. The maximum performance for both the atom transport and coherence is achieved in the non-interacting limit $(a_{s}\to 0)$. This conclusion stands in contrast to the interpretation of experimental results and many-body theories that propose the atomtronic transistor as a device for coherent matter wave emission precisely due to interatomic interactions \cite{PhysRevA.104.033311}. This framework suggests that coupling between an oscillating condensate in the gate well and the transistor modes, which is an inherently interaction driven mechanism, is responsible for the current gain and the coherence. Our analysis however finds no evidence for such a gain mechanism within the GPE model. Instead, we found that the interactions in the mean-field approximation acts as a detrimental factor that impedes the single particle resonances which serve as the primary source of coherence. Identifying which prediction is correct is an important topic for future theoretical work and experiements.
\section{Conclusion}
In this work, we have analyzed the atomtronic triple-well transistor using the Gross-Pitaevskii equation, one of the simplest mean-field models for studying interacting Bose-Einstein condensates near $T=0$. Within this framework, we found that the flow of atoms can be controlled by varying the bias potential in the source well, with transport occurring in resonant peaks as the source chemical potential aligns with the discrete energy levels of the gate. 

A primary goal of this study was to test the prediction of coherent matter-wave emission from the transistor and compare our findings with the theoretical framework proposed in  \cite{PhysRevA.104.033311}. Our simulations support the emission of a matter wave from the drain, which is coherent and oscillates at the characteristic single-particle frequency of the gate well. However, the underlying mechanism driving this coherence in our model appears to diverge significantly from previous theory.

Previous many-body treatment posits that transistor gain and coherence originate from a specific interaction-driven mechanism. In this model, a condensate forms in the gate as an oscillating displaced ground state (a truncated coherent state). This oscillating condensate actively couples to the transistor modes, broadening the energy range of incoming atoms that can tunnel to the drain. This coupling, driven by inter-particle interactions, is predicted to result in a substantial ``current gain'' and emit a ``classical coherent matterwave'' whose phase is locked to the gate's condensate oscillation. In this view, interaction is the essential engine for an enhanced, coherent output. In contrast, our mean-field simulations suggest that coherence in the drain is primarily governed by single-particle resonances. We observe coherent emission when the source chemical potential matches a discrete eigenenergy of the gate well, a process characteristic of standard resonant tunneling. Critically, our results show that the highest degree of coherence is achieved in the non-interacting or weakly interacting limit, and this coherence degrades as the interatomic interaction strength increases. We did not find any evidence of the coupling of transistor modes to a gate well condensate leading to a subsequent broadening of the tunneling width for the atoms.

The interaction between thermal and the condensate atoms, which is completely neglected in our work can qualitatively change the dynamics. Thermal atoms with sufficient energy can traverse the barriers to give rise to new physics that is not captured by the mean-field GPE. To include these finite temperature effects, extensions of the GPE such as the projected GPE \cite{PhysRevLett.87.160402}, Zaremba-Nikuni-Griffin (ZNG) \cite{Proukakis_2008, Griffin_Nikuni_Zaremba_2009} formalisms are necessary. These models include the interaction between the thermal and the condensate component in the atom ensemble and thus could provide better description of transistor dynamics as observed in experiments.
\section{Acknowledgements}
M.K. and S.Dowarah acknowledge support from the National Science Foundation (NSF) under Grant No. OSI-2228725. M.Du acknowledges support from the National Science Foundation (NSF) under Grant No. OSI-2228725. C.Z. acknowledges support from Air Force Office of Sci-
entific Research under Grant No. FA9550-20-1-0220 and
the National Science Foundation (NSF) under Grant No. PHY-
2409943, OSI-2228725, ECCS-2411394. S.Du and A.Z. acknowledge support from NSF (2228725 and 2500662) and DOE (DE-SC0022069). The authors thank the High Performance Computing facility at The University of Texas at Dallas (HPC@UTD) for providing computational resources.
\bibliography{apssamp}
\end{document}